\documentclass[11pt,pra,tightenlines]{revtex4}
\usepackage{graphicx,amsmath,amssymb}
\begin{document}
\title{Joint multipartite photon statistics by on/off detection}
\author{G. Brida$^1$, M. Genovese$^1$,  M. G. A. Paris $^2$, F.
Piacentini$^1$} \affiliation{$^1$ I.N.RI.M., Strada delle Cacce
91, 10135 Torino, Italia} \affiliation{$^2$ Dipartimento di Fisica
dell'Universit\`a di Milano, Italia}
\begin{abstract}
We demonstrate a method to reconstruct the joint photon
statistics of two or more modes of radiation using on/off
photodetection performed at different quantum efficiencies.
The two-mode case is discussed in details and experimental
results are presented for the bipartite states obtained
after a beam-splitter fed by a single photon state or a
thermal state.
\end{abstract}
\maketitle
The reconstruction of the joint photon distribution of two or more
correlated modes of radiation plays a crucial role in fundamental 
quantum optics \cite{MG} and finds relevant applications in 
quantum communication \cite{cav}, imaging \cite{lug} and spectroscopy 
\cite{spe}. Nevertheless, photodetectors suited for this purpose 
are currently not available, since the few existing examples 
\cite{riv} still suffer from limitations. On the other hand,
reconstruction by quantum tomography \cite{mun} is not
an easily implementable technique suited for a widespread use.
\par
Recently \cite{nos}, a maximum-likelihood (ML) method based on 
on/off detection performed at different quantum efficiencies
\cite{pco} has been developed and demonstrated for reconstructing 
the photon distribution of single-mode states. The results are 
reliable and accurate also for relatively low quantum efficiency 
of the detector. Since for many applications multipartite states 
are needed, in this letter we extend our previous results to this 
case as well. In particular, the bipartite case (easily extendable 
to multipartite) will be discussed in details. Examples of 
experimental reconstructions for bipartite state are 
presented to test and assess our method.
\par
The statistics of on/off detection performed with quantum
efficiency $\eta$ on a single-mode state $\rho$ is given 
by $p_{0\eta} = \sum_n A_{\eta n} \rho_n$ [$p_{1\eta} = 
1- p_{0\eta}$] where $\rho_n=\langle n |\varrho |n\rangle$ is 
the photon distribution (diagonal matrix elements) of the state,
and $A_{\eta n} = (1-\eta)^n$. By performing independent on/off
photodetection on two (spatially separated) modes of 
radiation, globally described by the two-mode density matrix 
$\varrho$, the joint on/off statistics is given by
\begin{align}
p_{00\eta} &= \sum_{nk} A_{\eta n} A_{\eta k} \varrho_{nk}\\
p_{01\eta} &= \sum_{nk} A_{\eta n} (1- A_{\eta k}) \varrho_{nk}\\ 
p_{10\eta} &= \sum_{nk} (1-A_{\eta n}) A_{\eta k} \varrho_{nk}\:,
\end{align}
and, of course, $p_{11\eta} = 1- p_{00\eta} - p_{10\eta} - 
p_{01\eta}$ where $\varrho_{nk}=\langle\langle nk | \varrho |nk
\rangle\rangle$ ($|nk\rangle\rangle = |n\rangle \otimes |k \rangle$) 
is the joint photon distribution of the two modes. Once the value 
of the quantum efficiency is known, the above equations provide a 
relation between the statistics of clicks and the actual statistics 
of photons. At a first sight this represents a scarce piece of 
information about the state under investigation. However, if the 
on/off statistics is collected for a suitably large set of efficiency 
values, then the information is enough to reconstruct the joint photon 
distribution of the bipartite state. We adopt the following strategy: 
by placing in front of the detector $K$ filters with different 
transmissions, we may perform the detection with $K$ different 
values $\eta_\nu$, $\nu=1...K$, ranging from $\eta_1=\eta_{\min}$ 
to a maximum value $\eta_K=\eta_{\max}$ equal to the nominal 
quantum efficiency of the detector. Upon writing 
$\mathbf{g}=(p_{00\eta_1},... ,p_{00\eta_K},
p_{01\eta_1},...,p_{01\eta_K},
p_{10\eta_1},...,p_{10\eta_K})$ and
$\mathbf{q}=(\varrho_{00},\varrho_{01},\varrho_{10},...)$,
according to 
$\varrho_{nk}\rightarrow q_p$ with $p=1+k + n (1+N)$, {\em i.e.}
$k=(p-1) \mod (1+N)$ and $n=(p-1-k)/(1+N)$ we can summarize 
the on/off statistics as 
\begin{equation}\label{con}
g_\mu = \sum_p B_{\mu p} q_p
\qquad\mu=1,..3K \quad p=1,..,(1+N)^2
\:,
\end{equation}
where we have introduced the matrix $\mathbf{B}$ 
\begin{align}
B_{\mu p} = \left\{
\begin{array}{ccl}
A_{\mu n} A_{\mu k} & & \mu=1,..,K \\
A_{\mu n} (1- A_{\mu k}) & & \mu=K+1,..,2K \\
(1 - A_{\mu n}) A_{\mu k} & & \mu=2K+1,..,3K
\end{array}
\right.
\end{align}
If the $\varrho_{nk}$'s are negligible for $n,k>N$ and  the
$\eta_\mu$'s are known, then Eq. (\ref{con}) represents a finite
statistical linear model for the positive unknown $q_p$. The
maximum-likelihood (ML)
solution of this LINPOS problem is well approximated by the
iterative algorithm\cite{EMalg}
\begin{align}
\label{iterazio}
q^{(i+1)}_p =&q^{(i)}_p \left(\sum_{\mu=1}^{3K}
B_{\mu p}\right)^{-1}
\sum_{\mu=1}^{3 K}
B_{\mu p}\:\frac{h_\mu}{g_\mu [\{q^{(i)}_p\}]}\:.
\end{align}
In Eq. (\ref{iterazio}) $q_p^{(i)}$ denotes the $p$-th element of
reconstructed  statistics at the $i$-th step,
$g_\mu[\{q^{(i)}_p\}]$  the theoretical on/off probabilities as
calculated from Eq. (\ref{con}) at the $i$-th step, whereas
$h_\mu$ are the measured frequency of the events with quantum efficiency
$\eta_\mu$, {\em i.e} 
$\mathbf{h}=(f_{00\eta_1},... ,f_{00\eta_K},
f_{01\eta_1},...,f_{01\eta_K},
f_{10\eta_1},...,f_{10\eta_K})$, 
with
$f_{ij\eta_\mu}=n_{ij\eta_\mu}/n_\mu$,
$n_\mu$ being the
total number of runs performed with $\eta=\eta_\mu$. \\ 
The convergence of the algorithm may be checked by the 
total error $$\epsilon_i = (3K)^{-1}\sum_\mu | h_\mu-g_\mu[\{q^{(i)}_p\}]|\:,$$
which measures the distance of the reconstructed statistics (of clicks) 
from the measured one: the algorithm is stopped when $\epsilon$ reaches its 
minimum, or goes below a certain threshold value.
\begin{figure}[h]
\includegraphics[width=0.35\textwidth]{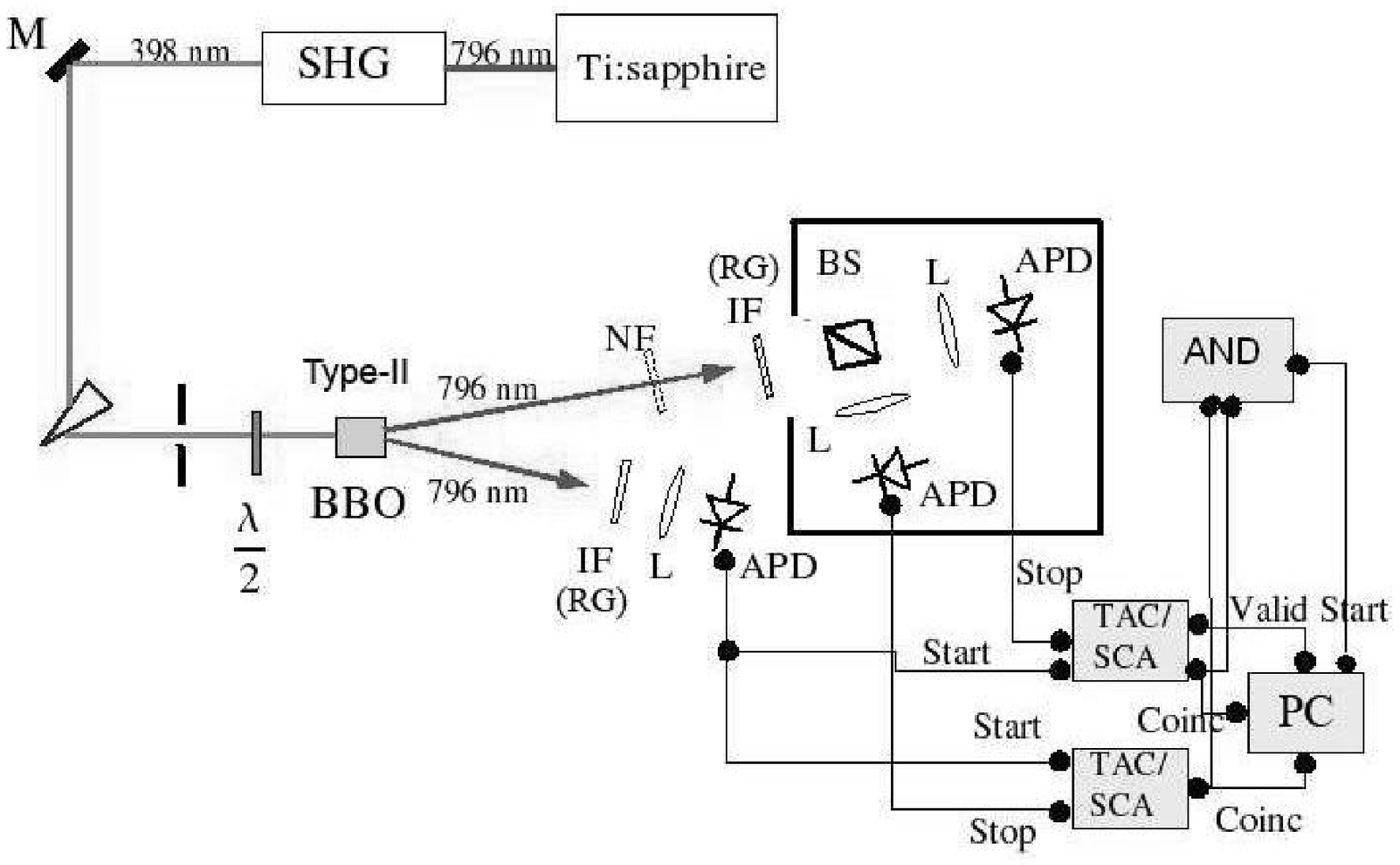}
\includegraphics[width=0.35\textwidth]{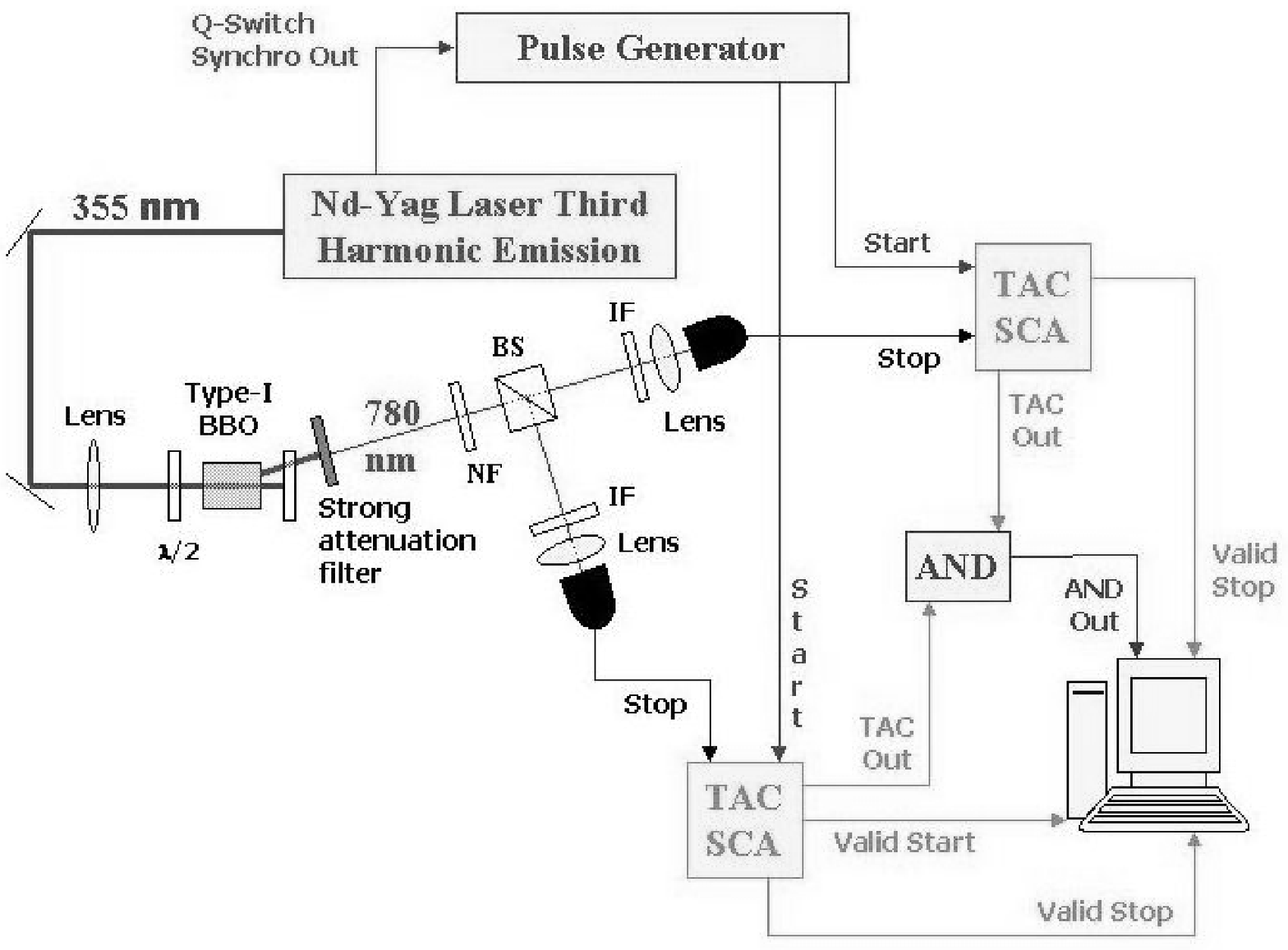}
\caption{Schematic diagrams of the experimental setups. On the 
left: setup to reconstruct the joint photon statistics of the
bipartite (entangled) states obtained by splitting a PDC-II heralded 
single-photon state by balanced and unbalanced beam-splitters. 
On the right: setup to reconstruct the joint photon 
statistics of a bipartite (classically correlated) state obtained by 
splitting a multithermal state by a balanced beam splitter. \label{f:setups}}
\end{figure}
\par
In order to test our method we have experimentally performed the 
reconstruction for different bipartite states. The schematic diagrams 
of the experimental setups are reported
in Fig. \ref{f:setups}.
As a first test, we use our method to reconstruct the bipartite 
(entangled) states obtained at the output of a beam splitter
fed by the heralded single-photon state generated by parametric 
down conversion (PDC). In our set-up a 0.2 W, 398 nm pulsed 
(with 200 fs pulses) laser beam, generated by second 
harmonic of a titanium - sapphire beam at 796 nm, pumps a type 
II BBO 5x5x1 mm crystal. Upon detecting a photon in a branch of the 
degenerate PDC emission one triggers the presence of the correlated 
photon in the other direction. The  heralded single-photon is then
impinged onto a beam splitter (BS) with unexcited second port, thus 
generating a bipartite entangled state of the form $|\psi\rangle\rangle 
= \sqrt{\tau} |01\rangle\rangle \pm \sqrt{1-\tau}|10
\rangle\rangle$, $\tau$ being the transmissivity. 
After the BS both arms are measured by an on/off detectors. 
All the detectors were APD silicon
photodetectors, whose quantum efficiency has been calibrated with the
traditional PDC scheme \cite{pdccs}. The proper set of quantum
efficiencies is  obtained by inserting before the BS several
Schott neutral filters (NF) of different transmittance, evaluated
by measuring the ratio between the counting rates with the filter 
inserted and without it. The data for the reconstruction have been 
taken using $K=34$ values of $\eta$ from $\eta_{\min}=0.015$ to 
$\eta_{\max}=0.325$. 
In correspondence of the detection of a photon in arm 1, a
coincidence window has been opened on both detectors on arm 2. This is
obtained by sending the output of the first detector as start to two
Time to Amplitude Converters (TAC) that receive the
detector signal as stop. The 20 ns window is set to not 
include spurious coincidences with PDC photons of the
following pulse (the repetition rate of the laser is 70 MHz). 
The TAC outputs are then addressed both to counters and to an AND 
logical gate for measuring coincidences between them. These outputs, 
together with one TAC Valid Start (giving us the total number 
of opened coincidence windows), allows to evaluate the frequencies 
on/off $h_\mu$ needed for reconstructing the joint photon statistics 
of the bipartite state. The background has been evaluated and 
subtracted by measuring the TAC and AND outputs out of the triggered 
window.
\par
\begin{figure}
\includegraphics[width=0.5\textwidth]{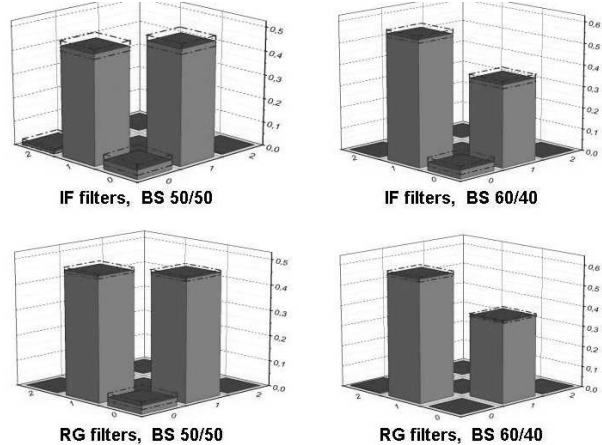}
\caption{\label{hera}Reconstruction of the joint photon distribution 
of a bipartite state: PDC heralded photon 
state split by a beam splitter (balanced and unbalanced). 
Dashed lines represent one standard deviation.}
\end{figure}

In order to verify the method in different cases we
considered 4 different alternatives given by the combination of a
balanced ($\tau=0.5$) or unbalanced ($\tau=0.4$) BS with
either large band, red glass filters (RG) with cut-off wave length
at 750 nm, or interference filters (IF), with peak wave length at
796 nm and a 10 nm FWHM.
The reconstructed statistics for these four situations are shown
in Fig. \ref{hera}. The uncertainties have been evaluated as 
described in Ref. \cite{nos}.
As it is apparent from Fig. \ref{hera} the reconstructed
state well corresponds to a  single photon in one of the modes.
Only the elements $\varrho_{01}$, $\varrho_{10}$ are different
(within uncertainties) from zero and their ratio is the value
expected by the ratio of the outputs ports of the BS (unity for
the balanced one, 2/3 for the unbalanced one).
As expected in this regime, no multi-photon component is observed:
i.e. $\varrho_{11}$, $\varrho_{20}$, $\varrho_{02}$, and so on are 
zero within uncertainties. The small uncertainties show that also 
less unbalanced BS would be distinguishable.
\par
As a second example we consider a single branch of PDC emission
without triggering, which corresponds to a multithermal state 
with number of modes of the order of $\sim 10^3$. A bipartite state is 
generated by impinging this signal onto a beam splitter with the 
second port unexcited. The output bipartite state is classically
correlated (not entangled, but not factorisable), with the two 
partial traces corresponding to multithermal states. The expected 
on/off statistics is given by 
\begin{align}
p_{00\eta}&= \mu^\mu (\mu+\eta N)^{-\mu}\\
p_{01\eta}&= \mu^\mu \left[ (\mu+\eta \tau N)^{-\mu} - (\mu+
\eta N)^{-\mu}\right] \\
p_{10\eta}&= \mu^\mu \left[
(\mu+\eta (1-\tau)N)^{-\mu} - (\mu+\eta N)^{-\mu}\right]
\end{align}
respectively, where $N$ is the average number of photons 
and $\mu$ the number of modes.
\begin{figure}[h]
\begin{tabular}{cc}
\includegraphics[width=0.3\textwidth]{Fig2_p1.ps} &
\includegraphics[width=0.3\textwidth]{Fig2_p2.ps} \\
\includegraphics[width=0.3\textwidth]{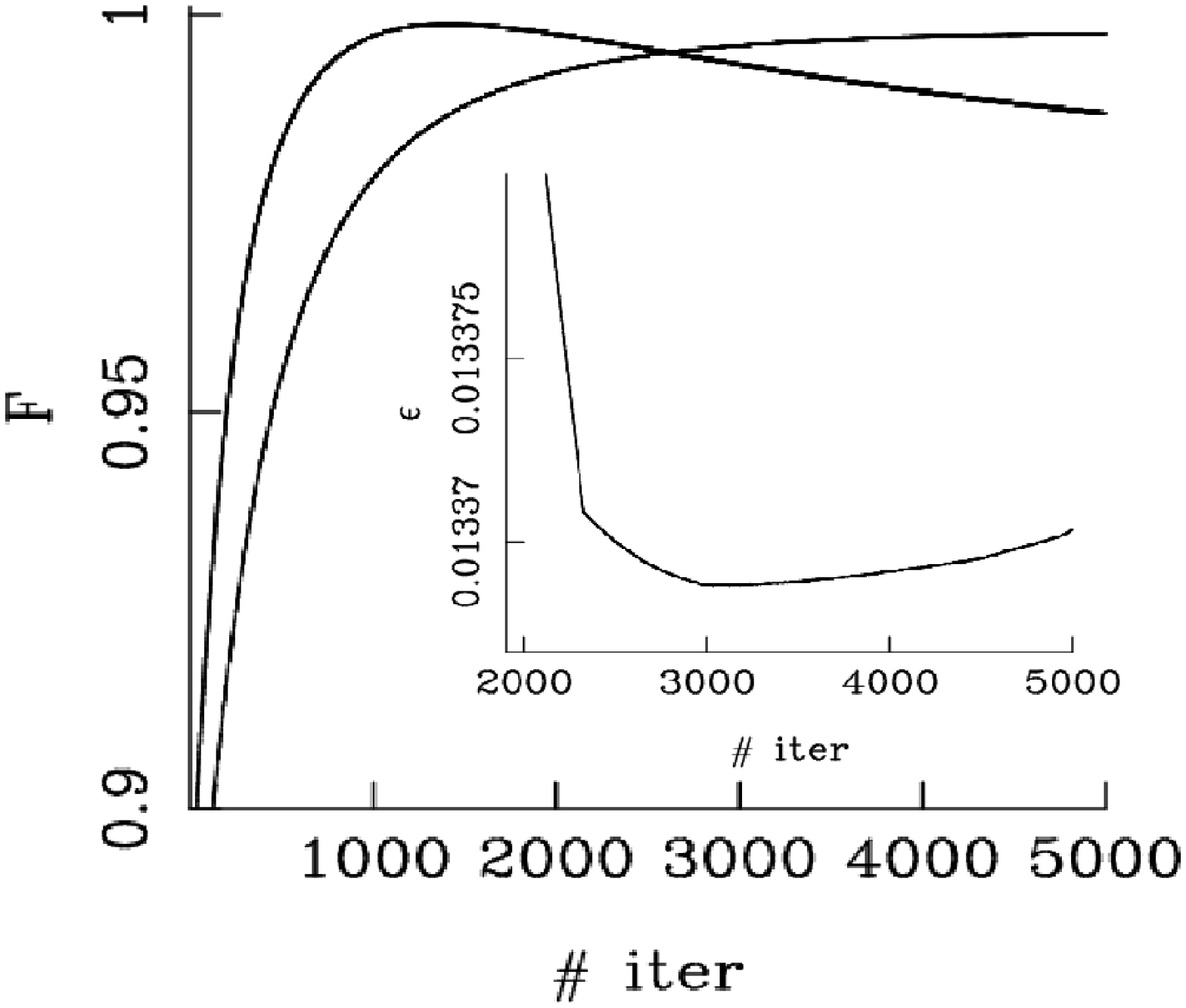} &
\includegraphics[width=0.3\textwidth]{Fig2_click.ps}
\end{tabular}
 \caption{\label{multi} Reconstruction of the joint photon distribution 
of a bipartite state: attenuated PDC multithermal distribution split by 
a balanced beam splitter.  The two histograms correspond to 
reconstructed distributions (dark gray) compared to multithermal 
ones. On the bottom left: fidelity $F$ of the reconstructed 
marginal distributions to multithermal ones as a function of the 
number of the iterations of the algorithm. The inset show the 
total error $\epsilon$. The last box show the measured frequencies 
$f_{00},f_{01},f_{10}$ as function of the quantum efficiency 
compared with the expected ones (multithermal, solid lines). }
\end{figure}
\par
In this case the state has been produced by pumping a 5x5x5 mm 
type I BBO crystal by a beam of a Q-switched triplicated (to 355
nm) Neodimium-Yag laser with pulses of 5 ns, power up to 200 mJ per
pulse and 10 Hz repetition rate.
Due to the very high power of the pump beam, a state with a large
number of photons is generated. We have
therefore attenuated (by 1 nm FWHM IF and neutral filters) the
multithermal state before splitting and detection.
Measurements at the different quantum efficiencies has been 
obtained by inserting (before the BS) Schott neutral filters, 
whose calibration has been obtained by measuring the power of 
a diode laser (at the same wave length of the used PDC emission) 
before and after them. The data for the reconstruction has been 
taken using $K=35$ values of $\eta$ from $\eta_{\min}=0.05$ to 
$\eta_{\max}=0.25$. 
The coincidence scheme has been realized by addressing two Q-switch
triggered pulses to two TAC modules as starts, and the detectors
outputs as stops. Then, having  set properly a 20 ns coincidence
window, we sent the two TAC outputs to an AND logic port, and the
Valid Stops to  counting modules (together with one TAC Valid
Start and the AND output).
The results of this reconstruction are shown in Fig. \ref{multi}.
Also in this case the comparison among theoretical expectations 
and reconstructed statistics is rather good. The fidelity 
$F=\sum_n \sqrt{\varrho_n \varrho_n^{mth}}$ of the reconstructed 
distribution to the expected multithermal $\{\varrho_n^{mth}\}$is larger than 
$99\%$ for both the marginals. Notice that the optimal number of iterations
({\em i.e.} leading to maximum average fidelity of the two marginals) 
corresponds to the minimum of $\epsilon$, thus confirming the 
good convergence properties of the algorithm.
\par
In conclusion, we demonstrated a method to reconstruct the 
joint photon statistics of two or more modes using on/off
photodetection. Experimental reconstruction have been presented 
for the bipartite states obtained after a beam-splitter fed 
by a single photon state or a thermal state. Our results clearly show 
that the ML reconstruction based on on/off detection can be successfully
applied to measure the joint photon statistics for multipartite systems.
\par $ $ \par 
This work has been supported by MIUR (FIRB RBAU01L5AZ-002 and
RBAU014CLC-002, PRIN 2005023443-002 and 2005024254-002), by
Regione Piemonte (E14), and by "San Paolo foundation".

\end{document}